\documentstyle[preprint,aps]{revtex}
\begin{document}
\setlength{\parindent}{0pt}
\title{A Renormalization-Group approach to the Coulomb Gap}
\author{S.R. Johnson$^{\dag,\ddag}$ \and D.E. Khmelnitskii$^{\dag,*}$}

\address{$\dag$ TCM Group, Cavendish Laboratory, Madingley Road,
Cambridge CB3 OHE \\ $*$ L.D. Landau Institute for Theoretical
Physics, Moscow, Russia \\ $\ddag${ \tt srj1001@phy.cam.ac.uk}}

\date{\today}
\maketitle
\begin{abstract}

The free energy of the Coulomb Gap problem is expanded as a set of
Feynman diagrams, using the standard diagrammatic methods of
perturbation theory. The gap in the one-particle density of states due
to long-ranged interactions corresponds to a renormalization of the
two-point vertex function. By collecting the leading order logarithmic
corrections we have derived the standard result for the density of
states in the critical dimension, \(d=1\).  This method, which is
shown to be identical to the approach of Thouless, Anderson and Palmer
\cite{TAP} to spin glasses, allows us to derive the strong-disorder
behaviour of the density of states. The use of the renormalization
group allows this derivation to be extended to all disorders, and the
use of an epsilon-expansion allows the method to be extended to
\(d=2\) and \(d=3\). We speculate that the renormalization group
equations can also be derived diagrammatically, allowing a simple
derivation of the crossover behaviour observed in the case of weak
disorder.

\end{abstract}

\input epsf
\vspace{1cm} 
\renewcommand\floatpagefraction{.9}
\renewcommand\topfraction{.9}
\renewcommand\bottomfraction{.9}
\renewcommand\textfraction{.1}

\section{Introduction}

The phenomenon of the Coulomb Gap has been known for 20 years
\cite{efros} and has been observed both in disordered semiconductors
below the Anderson transition \cite{efros2} and in the high magnetic
field Quantum Hall fluid \cite{hall}. In any system of localized
electrons, when electrostatic interactions are taken into account, a
soft gap appears in the density of states at the Fermi level. This
classical effect is the result of the interplay of disorder,
long-ranged interactions, and the discreteness of electric charge. The
Hamiltonian of the system \cite{efros,efros2} is

\begin{equation}
H[n_i] = \sum_i \phi_i n_i + e^2 \sum_{i \neq j}
\frac{(n_i-\frac{1}{2})(n_j-\frac{1}{2})}{r_{ij}}
\end{equation}

where the \(n_i\) are occupation numbers. The hamiltonian contains two
terms; the first dependant upon disorder and the second upon Coulomb
interactions. The \(\phi_i\) are uncorrelated energies representing
the effects of disorder, and macroscopic thermodynamic quantities must
be obtained by averaging over them according to the distribution:

\begin{equation}
\langle F \rangle = \langle F[\phi_i] \rangle =
\int_{-\infty}^{+\infty} \prod_i \frac{d\phi_i}{\sqrt{2\pi A^2}} \exp
(-\sum_i \frac{\phi_i^2}{2A^2}) F[\phi_i]
\label{avg} \end{equation}

There are two dimensionless parameters governing the problem. The one
describing the relative strengths of disorder and interactions is
\begin{equation}
\gamma = \frac{\varepsilon_0}{A}=\frac{e^2}{a A}
\end{equation}

where \(a\) is the lattice spacing, so that \(\varepsilon_0\) gives a
measure of the strength of nearest-neighbour interaction. We consider
the problem on a lattice, although because the system is dominated by
long-range behaviour over the order of the correlation length (\(r
\sim \frac{e^2}{T}\)) this should be irrelevant to the physics of the
problem. The standard Coulomb Gap problem in strong disorder
corresponds to \(\gamma \ll 1\), whilst \(\gamma \gg 1\) corresponds
to weak disorder. The dimensionless parameter describing thermal
effects is
\begin{equation}
\xi = \frac{\varepsilon_0}{T}=\frac{e^2}{a T}
\end{equation}

Because the distribution of the \(\phi_i\) is symmetric under \(\phi_i
\rightarrow -\phi_i\), the Hamiltonian is symmetric under \(\phi_i
\rightarrow -\phi_i, n_i \rightarrow 1-n_i\), so that the chemical
potential \(\mu=0\) and the total number of particles is fixed
\begin{equation}
\sum_i n_i = N/2
\end{equation}

We would like to derive the one-particle density of states (OPDOS),
\(g(\varepsilon)\), defined as the distribution function of the single
site energies

\begin{equation}
\varepsilon_i = \left(\frac{\partial H}{\partial n_i}\right)_{n_j} =
\phi_i + e^2 \sum_{j \neq i} \frac{(n_j-\frac{1}{2})}{r_{ij}}
\end{equation}

At nonzero temperatures the problem was treated by Mogilyanskii and
Raikh, \cite{mogil} and in particular it was shown that the
distribution of electrons over sites obeys Fermi-Dirac statistics

\begin{equation}
\langle n_i \rangle = n_{fd}(\varepsilon_i)= \frac{1}{1+\exp
\varepsilon_i/T}
\end{equation}

Once the OPDOS is known, we can derive many useful macroscopic
quantities such as screening law and conductivity.

All standard discussions of the Coulomb Gap focus on the strong
disorder problem, although we will also consider the case of weak
disorder. In this case we might expect a crystalline groundstate and a
hard gap. However, as noted by Efros \cite{efprl}, the freezing
temperature of a pure ionic crystal with no disorder is numerically
smaller by about two orders of magnitude than \(\varepsilon_0\). Thus
there is a very wide range of temperatures between this freezing
temperature and \(\varepsilon_0\) for which the weakly disordered
Coulomb Gap system behaves like a strongly correlated classical fluid.

In section \ref{per}, we demonstrate that the Coulomb Gap partition
function can be expanded as a series of Feynman diagrams. The general
philosophy is to expand the partition function \(\ln Z[\phi_i]\) as a
series of diagrams and only at the last moment to perform the
averaging over disorder (\ref{avg}). This method is identical to the
method used by Thouless, Anderson, and Palmer (TAP) \cite{TAP} to
study the spin glass problem; we discuss this in section \ref{rel}. In
the next section \ref{im}, we give a simple argument for the lack of
any low-temperature phase transition in the critical dimensionality
\(d=1\). This means that the derivation of the leading approximation
to the one particle density of states for the Coulomb Gap problem
given in section \ref{der} does indeed sum the most important set of
diagrams, and that any divergences in other sets of diagrams must
cancel. The standard results in \(d=2\) and \(d=3\) can be obtained by
means of an epsilon-expansion in \(\epsilon=d-1\), as explained in
section \ref{rgt}. Finally, we speculate that renormalization group
theory arguments can allow a derivation of the weak-disorder limit in
which the standard results for the density of states break down.

\section{Diagrammatic perturbation theory for the Coulomb Gap} \label{per}

The density of states is calculated by means of a diagrammatic
perturbation theory. Polyakov \cite{poly} shows how to convert
classical Ising-like models intoa continuous field problem, which can
be expanded into the Feynman diagrams well-known from quantum field
theory, and we use a slightly-modified version to take account of
disorder.

The partition function is rewritten using the identity

\begin{eqnarray}
Z[\phi_i] &=& \sum_{[n_i]} \exp \left( -\frac{H[n_i,\phi_i]}{T}\right)
\nonumber \\ &=& \sum_{[n_i]}\exp\frac{1}{T}\left( \sum_i \phi_i
(n_i-\frac{1}{2}) + e^2 \sum_{i \neq j}
\frac{(n_i-\frac{1}{2})(n_j-\frac{1}{2})}{r_{ij}}\right) \label{z} \\
&=& \sum_{[n_i]}\exp\left( \sum_i \frac{\phi_i}{T}
\left. \frac{d}{d\chi_i}\right|_0 + e^2 \sum_{i \neq j}
\frac{T}{r_{ij}}\left. \frac{d}{d\chi_i}\right|_0
\left. \frac{d}{d\chi_j}\right|_0\right)\exp \sum_i \frac{\chi_i
(n_i-\frac{1}{2})}{T} \end{eqnarray}

leaving the result in a form in which the summation over \([n_i=0,1]\)
can be easily performed, using the identities:
\begin{eqnarray}
\sum_{[n_i]} \exp \sum_i \frac{\chi_i(n_i-\frac{1}{2})}{T} &=& \prod_i
\cosh \frac{\chi_i}{2T} = \exp \sum_i \ln \cosh \frac{\chi_i}{2T}
\nonumber \\ \exp \sum_i \frac{\phi_i}{T} \frac{d}{d\chi_i} \exp
\sum_i \ln \cosh \frac{\chi_i}{2T} &=& \exp \sum_i \ln \cosh
\frac{\chi_i+\phi_i}{2T}
\end{eqnarray}

to yield the formally exact result:
\begin{equation}
Z[\phi_i] = \exp( \sum_{i \neq j}
\frac{-Te^2}{r_{ij}}\left. \frac{d}{d\chi_i}\right|_0
\left. \frac{d}{d\chi_j}\right|_0)\exp -\sum_i \ln \cosh
(\frac{\chi_i+\phi_i}{2T}) \label{diag}
\end{equation}
Our general approach will now be to perturbatively expand \(\ln
Z[\phi_i]\) as a series of connected Feynman diagrams, similar to the
'locator' perturbation series of Anderson \cite{anderson}, and only at
the very last stage to perform the averaging over \(\phi_i\). This
corresponds to the approach of Thouless, Anderson and Palmer (TAP)
\cite{TAP} to spin glasses.

The diagram rules for the series expansion of \(\ln Z\) are as
follows, where \(n_{fd}(\phi)\) is the Fermi-Dirac function:

\begin{itemize}
\item{Each vertex \(i\) with \(n_i\) lines coming out of it gives a factor 
of 
\begin{equation}
-(T \frac{d}{d\phi_i})^{n_i}\ln \cosh \frac{\phi_i}{2T} = 
(T \frac{d}{d\phi_i})^{n_i-1} (n_{fd}(\phi_i)-\frac{1}{2}) \nonumber
\end{equation}}
\item{Each line from \(i\) to \(j\) gives a factor of 
\begin{equation}
\frac{-e^2}{r_{ij}T} \nonumber
\end{equation}}
\item{Symmetry factor \(1/|G|\), where \(|G|\) is a combinatorial
factor equal to the order of the symmetry group of the diagram. This
standard factor \cite{itz} is well known from statistical and quantum
field theories.}
\end{itemize}

When it comes to averaging over disorder, we note that the
disorder-averaged bare n-point vertex function is:

\begin{equation}
\int_{-\infty}^{\infty} g_0(\phi)d\phi \left(T \frac{d}{d\phi}
\right)^n \ln\cosh \frac{\phi}{2T}
\end{equation}

Integrating this by parts, we can see that it equals

\begin{equation}
\int_{-\infty}^{\infty} T dx \left(\frac{d}{dx} \right)^{n-2} g(T x)
 \left( \frac{d}{dx} \right)^2 \ln \cosh x \approx \pi T^{n-1}
 \left(\frac{d}{dx} \right)_{x=0}^{n-2} g(x)
\end{equation}

so that for all non-pathological distributions of disorder (such as
for example Gaussian (\ref{avg})) the bare two-point vertex function
is

\begin{equation}
\lambda_2 = \frac{\pi T a^d}{A} 
\end{equation}

(where \(d\) is the dimensionality of space) and higher vertex
functions disappear like powers of \(\frac{T}{A}\),

\begin{equation}
\lambda_n \propto (\frac{T}{A})^{n-1}
\end{equation}

so that they are negligible to our calculations to leading logarithmic
order. In the case of a 'top-hat' distribution of local site energies,
as used in numerical simulations, \(\lambda_n\) disappear like \(\exp
-\frac{A}{T}\) for \(n>2\). Thus to leading order, only two-point
vertices survive after averaging. Note that the bare two-point vertex
function is proportional to the non-interacting density of states
\(g_0\), and that the renormalized two-point vertex function is given,
after averaging, by

\newcommand{\sech}{\mbox{sech}}
\begin{equation}
\langle  \sech^2\frac{\varepsilon_i}{2T} \rangle = T a^d g \label{2pt}
\end{equation}
where \(g\) is the density of states at the Fermi level as a function
of temperature.

\section{The relationship between the Coulomb Gap and the TAP treatment 
of spin glasses} \label{rel}

The idea of performing a diagrammatic expansion of \(\ln Z\) and of
only performing the averages over disorder at the last moment is
identical to the TAP theory of spin glasses \cite{TAP}. The Coulomb
Gap problem is in some senses more simple than the spin glass problem,
because for \(A \gg T\) only two-point vertices survive after
averaging. The only reason that the field theory is not a trivial
\(\phi^2\) theory is the possibility that more than one vertex could
represent the same site, so that if for example \(i=k\), \(\phi_i\)
and \(\phi_k\) will no longer be independant. We represent this case
by Feynman diagrams including 'connections' (see figure
\ref{connect}). As in the diagrams for the conductivity of a dirty
metal \cite{abrikosov}, it is these connections which make the theory
non-trivial, in that there are one-particle irreducible diagrams
\cite{itz} more complex than a single bare two-point vertex.

In the next section we will show that a certain set of Feynman
diagrams (see figure \ref{paper}) which we will call the 'maximally
crossed' diagrams, are larger by \(O(\frac{A}{\varepsilon_0})\) than
any others, and thus dominate the expression for the density of
states.

This behaviour is identical to that of spin glasses, for which a
certain set of diagrams is larger than any others by \(O(z)\) where
\(z\), the coordination number, is assumed large. However in spin
glasses, the sum of a set of loop diagrams incorporating four-point
vertices, diverges at the spin glass temperature, signalling the onset
of replica symmetry breaking (RSB). However the TAP approach allows
expansions around frozen-in spins, and the method is 'fail-safe' in
the sense that if we are expanding about a state which is not a global
minimum of energy, we will get negative or zero eigenvectors of the
response matrix, and a negative or infinite susceptibility. The
low-temperature state is marginally stable \cite{bray} (meaning that
the eigenvalues of the stability matrix are distributed all the way
down to zero) but the TAP method holds up under these
circumstances. The marginal stability of the system is precisely due
to the exponentially large number of metastable states.

In section \ref{im} we note that there exists a simple physical
argument why no such phase transition can occur at low temperature in
the Coulomb Gap problem in \(d=1\). Thus any sets of diagrams whose
sums do diverge at low temperatures must cancel with other such
diagrams. In \(d=2, d=3\) the question of whether there exists a glass
transition is inconclusive, although this does not affect our
arguments.

Anderson \cite{anderson2} demonstrates a derivation of the
low-temperature distribution of effective local fields in the spin
glass problem which is exactly analogous to the standard derivation of
the density of states in the Coulomb Gap \cite{efros2}. However
because of the onset of RSB at low temperatures, a rather intricate
method due to Bray and Moore \cite{bray} is necessary in order to
perform a perturbation theory expansion around the non-trivial
ground-state of the spin glass, although their method is a locator
perturbation theory, essentially the same as our calculation in this
paper. Bray and Moore define a Greens function, which when averaged
over disorder, gives exactly (\ref{2pt}).

The Coulomb Gap is effectively a variant of the random-field
ferromagnet with long-ranged interactions. Bray \cite{bray2} confirms
that whilst the random-field ferromagnet displays low-temperature
anomalies in the specific heat (again due to marginal stability), it
does not have RSB.

\section{Absence of a low-temperature phase transition in \(d=1\)} \label{im}

As mentioned in the previous section, we are performing a perturbation
expansion around a trivial groundstate. Thus it is essential that we
sum the diagrams at a temperature higher than that of any phase
transition, and we will show that in \(d=1\) there is no such
transition.

Consider a 1-d Coulomb Gap system with small but finite disorder,
\(\gamma \gg 1\), at \(T=0\). The system will form into domains of
perfect ionic crystals separated by domain walls. Suppose that the
density of domain walls is \(N\), so that the mean number of sites in
a single domain is \(N^{-1}\): the limit \(\gamma \gg 1\) corresponds
to \(N \ll 1\). Then it is possible to make a simple argument similar
to that of Imry and Ma \cite{imry}, showing that the free energy has a
minimum for a finite domain wall density, of order
\begin{equation}
N_c \sim \frac{A^2}{\varepsilon_0^2}
\end{equation}

and so any disorder, no matter how small, destroys long range order in
\(d=1\). Thus in this critical dimensionality there is no low
temperature phase transition, and we are safe in the knowledge that
our perturbation theory is around a stable minimum.

Vojta \cite{vojta} shows that within the uncontrolled approximation of
the spherical model, the Coulomb Gap system has a low-temperature
transition to a phase with long-range order in \(d>4\), whereas in \(d
\le 4\) there is no such transition. Monte Carlo simulations are
inconclusive on the question of whether there is a phase transition in
\(d=2, d=3\) \cite{pollak,davies}. Studies of a related system, the
Coulomb Glass show a phase transition in \(d=3\) \cite{grannan} but
not in \(d=2\) \cite{xue}. All numerical simulations show clearly that
the Coulomb Gap is well-developed at temperatures well above the
suspected glass transition. This gives us a justification for
disregarding the glass transition and performing an expansion around
the trivial groundstate.

\section{Deriving the density of states within perturbation theory} 
\label{der}

We discuss first the critical dimensionality \(d=1\), because the
divergences encountered here are logarithmic and hence tractable to
the renormalization group. We mention later how the results are
modified for the more severe divergences in \(d=2, d=3\). We aim to
derive the results of Raikh and Efros \cite{raikh} for the Coulomb gap
in strong disorder
\begin{equation}
g = g_\infty =\frac{g_0}{1+e^2 g_0 \ln (\varepsilon_0/T)} \label{strong}
\end{equation}

and to show how these are modified in weak disorder. In d=2, where the
strong disorder density of states is given by
\begin{equation}
g_\infty=\frac{2T}{\pi e^4}
\end{equation}

Pikus and Efros \cite{pikus} observed a crossover behaviour for the
weak disorder problem, with a density of states given by

\begin{equation}
g \approx \frac{A}{\varepsilon_0} g_\infty + \frac{1}{e^2} \exp (-
\alpha \varepsilon_0/T) \label{weak}
\end{equation}

We note that heuristically, we expect the low-disorder density of
states to behave like the Boltzmann function \(\exp -\frac{\alpha
\varepsilon_0}{T}\) where the typical energy of excitations is
\(\alpha \varepsilon_0\).

We have already discussed how the diagrammatic perturbation expansion
of \(g\) can be obtained. The 'maximally crossed' diagrams (see figure
\ref{paper} for maximally crossed diagrams to order \(g_0^2\) ) are
larger by \(O(\frac{\varepsilon_0}{A})\) than any diagrams containing
unpaired vertices. Before averaging over disorder these diagrams give
\begin{eqnarray}
\sech^2 (\frac{\phi_i}{2T}) &=& \sech^2 (\frac{\varepsilon_i}{2T}) -
\sech^4 (\frac{\varepsilon_i}{2T}) \sech^2 (\frac{\varepsilon_j}{2T})
\left(- \frac{e^2}{r_{ij}T} \right)^2 \nonumber \\ &-& \sech^4
(\frac{\varepsilon_i}{2T}) \sech^4 (\frac{\varepsilon_j}{2T}) \left(-
\frac{e^2}{r_{ij}T} \right)^3 - \sech^6 (\frac{\varepsilon_i}{2T})
\sech^4 (\frac{\varepsilon_j}{2T}) \left( -\frac{e^2}{r_{ij}T}
\right)^4 +\cdots
\end{eqnarray}

We first perform the spatial integration, then by inspection sum the
series. As the final step of the calculation, we can now perform the
average over disorder (\ref{avg}) to give, using (\ref{2pt}), and to
logarithmic accuracy,

\begin{eqnarray}
g &\approx& g_0 - \frac{e^2 g_0^2}{2} \ln (1
+\frac{\varepsilon_0^2}{T^2}) + \cdots \nonumber \\ &\approx& g_0 -
e^2 g_0^2 \ln \frac{\varepsilon_0}{T} \label{as}
\end{eqnarray}

As we would expect, there is the same logarithmic behaviour for a
\(\frac{1}{r^d}\) potential in \(d\) spatial dimensions.  For the
Coulomb Gap in \(d=1\) the divergences are mild enough that we can
collect together the leading order logarithmic terms depicted in
figure \ref{paper3} to give an expansion for \(g\) in terms of \(g_0\)

\begin{equation}
g \approx g_0 - e^2 g_0^2 \ln \frac{\varepsilon_0}{T} + e^4 g_0^2
\ln^2 \frac{\varepsilon_0}{T} - \cdots
\end{equation}

so that the strong disorder case can be derived as a resummation of
all the leading order logarithmic terms. A more satisfactory mechanism
for collecting together all terms of the same magnitude is to use the
renormalization group equations, and instead of considering a
perturbation expansion for \(g\), to derive an expansion the
beta-function. This is discussed in the next section.

\section{Renormalization group theory} \label{rgt}

We begin the renormalization group treatment by noting that we wish to
express the renormalization group equations in terms of dimensionless
quantities, such as the partition function Z (\ref{z}), and
\begin{equation}
\Gamma=e^2 a^{d-1} g 
\end{equation}

which is the probability that any site is within the gap. We will
express \(\Gamma\) in terms of \(\xi\), where
\begin{equation}
\xi = \frac{r_T}{a} =\frac{\varepsilon_0}{T}
\end{equation}
where the correlation length associated with a temperature T is \cite{hunt}

\begin{equation}
r_T=\frac{e^2}{T} \label{corr}
\end{equation}

The beta-function is

\begin{equation}
\beta(\Gamma) = \frac{d \ln \Gamma}{d \ln \xi} = \beta_0+\beta_1
\Gamma+\beta_2 \Gamma^2+\cdots \label{beta}
\end{equation}

For the case of strong disorder, solving the following equation for
the beta-function:

\begin{equation}
\beta(\Gamma) = -\Gamma \label{crossover0}
\end{equation}

together with the boundary condition \(\Gamma \rightarrow \Gamma_0\)
as \(\xi \rightarrow 0\) gives

\begin{eqnarray}
\Gamma &=& \frac{\Gamma_0}{1+\Gamma_0 \ln \xi} \nonumber \\ g_T(0) &=&
\frac{g_0}{1+e^2 g_0 \ln \frac{\varepsilon_0}{T}} \approx \frac{1}{e^2
\ln \frac{\varepsilon_0}{T}}
\end{eqnarray}

which is exactly the result of Raikh and Efros for the density of
states in \(d=1\) with strong disorder (\ref{strong}). In other
dimensionalities, an epsilon-expansion gives the beta-function to be

\begin{equation}
\beta(\Gamma) = -(d-1)-\Gamma 
\end{equation}

which can again be solved with the boundary condition \(\Gamma
\rightarrow \Gamma_0\) as \(\xi \rightarrow 0\) to give

\begin{eqnarray}
\Gamma &\propto& \xi^{-(d-1)} \nonumber \\
g_T(0) &\propto& \frac{T^{d-1}}{e^2d}
\end{eqnarray}

which is the standard result of Efros and Shklovskii
\cite{efros2}. Thus we reach our main conclusion, that the TAP method
for solving the spin-glass problem is an identical approximation to
the Efros-Shklovskii method for obtaining the density of states of the
Coulomb Gap.

We note that the \(d-1\) term can be derived using hyperscaling. The
system has a critical point which has been moved to \(T=0\), and the
correlation length (\ref{corr}) implies a critical index \(\nu =1\),
so by hyperscaling, the heat capacity must be proportional to
\(T^{d-2}\) and the density of states to \(T^{d-1}\).
 
In the limit of weak disorder \(\gamma \gg 1\), the standard results
for the density of states break down \cite{pikus}, and in this section
we explain how to use the renormalization group equation to derive the
behaviour in this regime.  The renormalization which we are performing
corresponds to a change in the disorder, \(\gamma\) of the system. As
noted in section \ref{der}, we expect a Boltzmann-like behaviour for
the density of states at low disorder.

\begin{eqnarray}
\Gamma &=& \Gamma_* \exp \left( - \frac{\varepsilon_0}{T} \right)
\nonumber \\ \ln \Gamma/\Gamma_* &=& -\xi \nonumber \\ \beta(\Gamma)
&=& \ln \Gamma/\Gamma_*
\end{eqnarray}

where \(\Gamma_*\) is an order unity constant. This beta-function is a
result which we expect to be universal at low disorder. By solving
equation (\ref{beta}) for the modified beta-function which crosses
over between the weak-disorder and strong-disorder limits,

\begin{eqnarray}
\beta(\Gamma) &\approx& -(d-1)- \Gamma \; \; (\Gamma \ll 1)
\label{crossover1} \\ \beta(\Gamma) &\approx& \ln \Gamma/\Gamma_* \;
\; (\Gamma \sim \Gamma_*) \label{crossover2}
\end{eqnarray}

with again the same boundary conditions \(\Gamma \rightarrow
\Gamma_0\) as \(\xi \rightarrow 0\), we obtain a density of states
with a crossover behaviour which in \(d=2\) coincides with that
(\ref{weak}) observed in the numerical simulations by Efros and Pikus
\cite{pikus}.

All the information in equations
(\ref{crossover0},\ref{crossover1},\ref{crossover2}) can be summarized
in figure \ref{graph}.Any glassy phase transition or crystallization
such as those discussed in section \ref{im} would constitute a
breaking of universality in the region \(\ln \Gamma \sim 0\). However
even for infinitesimal disorder, the site energies will have a spread
of order \(\varepsilon_0\) and \(\Gamma\) will be of order unity so
that this region is of little physical significance.

As noted earlier, there is a crossover between the strong-disorder
(Efros-Shklovskii) and weak-disorder (Boltzmann) behaviour. In \(d=1\)
this occurs at a value of

\begin{equation}
\Gamma = \Gamma_c = a +b \ln \Gamma_*
\end{equation}

with \(a,b\) order unity constants, whilst in \(d=2\) the modified
renormalization group equations give exactly the DOS observed in
\cite{pikus}.

The corrections to hyperscaling in the RHS of (\ref{beta}) can be
obtained as a systematic perturbation series by using the identity
\begin{equation}
\frac{d \langle \ln Z \rangle}{d \ln T} = \left\langle -\sum_i \phi_i
\frac{d \ln Z[\phi_i]}{d \phi_i}\right\rangle - \left\langle e^2
\frac{d \ln Z[\phi_i]}{d e^2}\right\rangle
\end{equation} 

which follows from (\ref{diag}). Thus we can develop a perturbation
theory for \(\beta\), the first non-zero term of which (see figure
\ref{connect2}) corresponds to the standard Coulomb gap result
(\ref{crossover0}, \ref{crossover1}). We believe that by evaluating
further terms in the series that we will obtain a series which
interpolates smoothly between (\ref{crossover1}) and
(\ref{crossover2}).

\section{Conclusions}

The essence of our paper is in linking two separate pieces of physics,
the spin glass problem, and that of the Coulomb Gap. We have
demonstrated that the standard results for the density of states in
the Coulomb Gap can be derived by a method identical to the TAP method
in spin glass theory.The Coulomb Gap density of states plays the part
of the renormalized two-point vertex function in an effective field
theory. In \(d=1\) we can perform the renormalization directly by
summing a dominant set of diagrams to obtain the strong disorder limit
of the Coulomb Gap. The results in \(d=1\) suggest the use of the
Renormalization Group to derive this result in any dimensionality, and
to extend the result into the case of weak disorder. It is shown how a
diagrammatic expansion for the \(\beta\)-function can be used to
derive the renormalization group equations. Thus we have demonstrated
that our formalism allows an understanding of the cross-over between
the strongly disordered Coulomb Gap problem and systems with weak
disorder.

\section{Acknowledgements}
We acknowledge useful discussions with Prof. A.J. Bray and
Prof. M.A. Moore. SRJ is supported by an EPSRC studentship.

\newpage
\vspace{1cm}
\begin{figure}[h]
\caption{An example of a Feynman diagram containing a connection; this
represents the possibility \(i=k\)}
\label{connect}
\end{figure}
\vspace{1cm}
\vspace{1cm}
\begin{figure}[p]
\caption{Diagrams giving to the standard Coulomb Gap density of
states; these 'maximally crossed' diagrams are larger by
\(O(\frac{A}{\varepsilon_0})\) than all others.}
\label{paper}
\end{figure}
\vspace{1cm}

\vspace{1cm}
\begin{figure}
\caption{The Dyson equation giving all the leading logarithm terms in
the density of states}
\label{paper3}
\end{figure}
\vspace{1cm}

\vspace{1cm}
\begin{figure}[p]
\epsfxsize=13cm
\caption{The scaling beta-function \(d\ln \Gamma / d \ln \xi\) plotted
as a function of \(\ln \Gamma\)}
\label{graph}
\end{figure}

\vspace{1cm}
\begin{figure}[p]
\caption{The first non-zero term in the perturbative expansion of the
beta-function. The wavy line represents a \(\phi d/d\phi\) term}
\label{connect2}
\end{figure}
\vspace{1cm}

\end{document}